# Assessment of Different Health Index Aggregation Techniques for Electric Utilities

Swasti R. Khuntia,
TenneT
Arnhem, The Netherlands
Swasti.Khuntia@tennet.eu

Bjoern Heling
TenneT
Bayreuth, Germany
Bjoern.Heling@tennet.eu

***Abstract*—The Health Index is a value (in terms of color or score) which describes the technical condition of an asset. Using the Health Index of various assets, the so called aggregated Health Index of a system can be calculated. For electric utilities, the different levels of aggregation can be at asset type (primary, secondary, etc.) or at bay or at substation level. This paper assesses the feasibility and methodology of aggregating Health Index scores with the help of different techniques. For validation, real values from five substations of the Dutch transmission system operator are used. The different aggregation techniques are compared by showing the effectiveness of aggregated methodology both in terms of quantity and quality. The aggregated Health Index score of all assets in a bay are integrated into bay Health Index.**

## I. INTRODUCTION

Asset management is defined as the process of maximizing the return on investment of asset over its entire life cycle by maximizing performance and minimizing costs (both capital expenditure and operational expenditure) at a given risk level [1]. In general, physical assets degrade over time and thus timely investments need to be planned and executed to keep a "healthy" asset portfolio. Asset managers are responsible for these physical assets and have a tough job of making an optimal refurbishment or replacement decision. They plan and prescribe relevant maintenance guidelines to study options that maximise the value of an asset as it approaches the end of its useful life. Maintenance used to be time-based or condition-based as prescribed by the manufacturer. In recent times, due to digitization, many utilities are adapting data driven asset management processes to streamline their budgeting process (i.e., capital expenses or CapEx on new assets and operational expenses or OpEx on existing assets) and optimize the available resources [2]. As a result, large amount of data (e.g., condition indicators) is collected where the aim is to better track and monitor the assets. To help in judicious decision making, this large amount of data needs to be transformed into knowledge. Although many utilities excel at data collection, but in reality few manage to use the data in a meaningful way. One way of translating such collected data to a key performance indicator (KPI) is through the Health Index (HI). It is an integral feature of complete asset management process. For an asset, at any given time, HI helps in understanding its health status, i.e., if it is in good condition or needs attention in terms of refurbishment or replacement. The asset HI is defined as a predictive analytics-based asset score that explains the asset condition and the likely performance based on characteristics such as condition risk. It is used to compare asset health and as a foundation for maintenance and replacement strategies by electric utilities [3]. At present, within TenneT B.V. (the Dutch TSO), the HI is calculated via three main parameters; expected lifespan based on assumed failure statistics of a population, the actual age based on the year of construction and the aging of individual component, expressed in a condition factor based on asset condition, of the component. It is illustrated in Fig. 1.

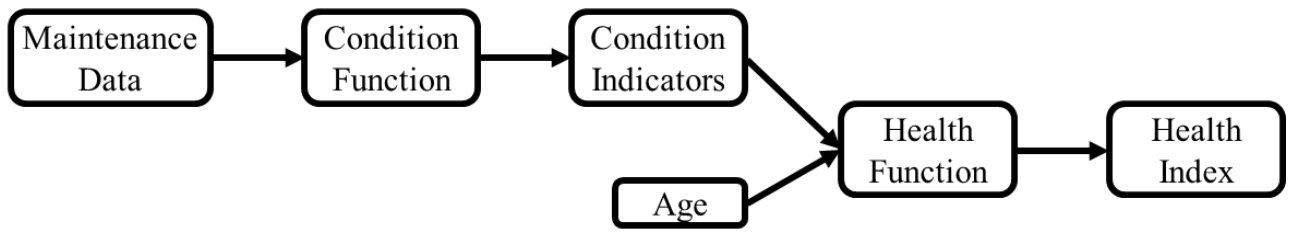

Figure 1. Asset HI definition at TenneT [4]

While the term "health" refers to the state of an asset and its proximity to the end of its useful life, such a definition of HI can be translated to other levels of aggregation within a substation, i.e., bay level or substation or asset type level. An example is a recent study on electrical tower HI calculation. Ref. [5] proposed a methodology where HI data was aggregated on tower level and a single HI value of each tower was generated on the basis of detected defects. In this work, we define aggregated HI as a practical method to quantify the general health at various levels of aggregation within a substation. The configuration of a substation is composed of multiple subsystems (called bay) and each subsystem is composed of multiple assets. As shown in Fig. 2, it can be visualized as a 3-layer system where each layer is associated with some decision making. It may be used as a tool to manage assets, to identify capital investment needs, to plan operational expenses (e.g., maintenance programs) and risk calculation. This might include partial asset replacement such as refurbishment to extend useful life, or an indefinite ongoing patch-and-continue programme, perhaps involving suppliers to provide necessary parts/services or similar activity [6]. In a nutshell, this information will help the asset managers to prioritize through potentially thousands of assets (or numbers of bays) and/or enable them to focus attention on the assets (or bays) that are in the worst condition. It is noteworthy to mention that the increasing level of aggregation is inversely proportional to the increasing level of details. The rest of this paper is organized as follows. Section II explains the concept of aggregation. Section III discusses and compares the different aggregation techniques. A comparative case study based on TenneT data is presented in section IV. Finally, section V summarizes this work.

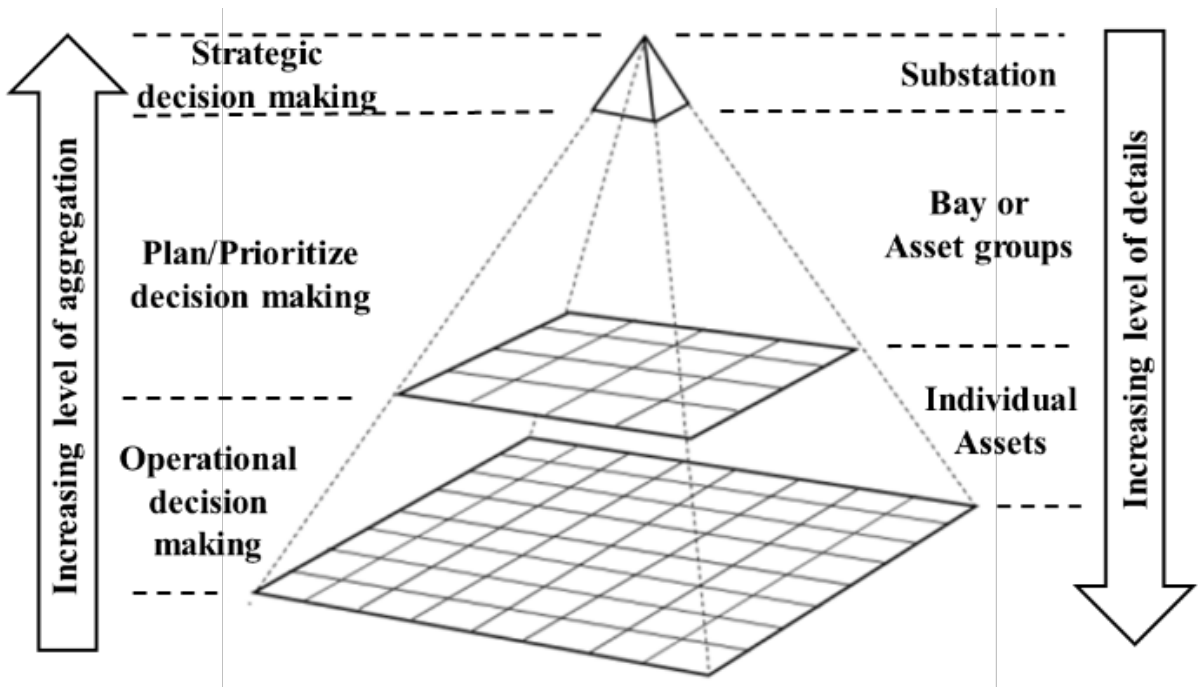

Figure 2. Aggregation layer for substation with decision making

## II. Aggregation: Theory and Practice in Substation

### *A. How to define aggregation for Health Index?*

As defined in previous section, asset HI is an aggregated way of measuring the overall health of an asset. It works on the principle that a list of asset attributes (such as condition indicators) are fed into the index calculation in some sort of aggregation method. By definition, aggregation is a process synthesizing in a global or aggregated value the information coming from various level of data sources. When the various data sources are combined, the aggregated value must satisfy and represent some preferences of the selected group of individuals and thus helping in decision making at different levels. Similar to asset HI, aggregated HI can be defined for any desired level which will be composed of specific data parameters at that desired level. Technically, aggregated HI is moving one level above the asset level and it can be summarized into a number range like 0-5 or 1-10 or vice versa. For e.g., a HI of range 1-10 can be interpreted as: 1 is the new health and 10 is the health condition when it could fail at any moment or vice versa. Literature study reveals few studies on HI aggregation theories and/or aggregation levels. One such study is ref [7] where the authors studied for the high voltage substation in distribution system using analytical hierarchy process. Another recent study [5] explains the authors' viewpoint on HI aggregation for electrical towers. A similar study aimed at parameter identification for predictive maintenance [8].

The critical objectives in the formulation of an aggregated HI can be adapted from the work [9]:

- The index should be indicative of the suitability of different assets at the chosen hierarchical level for a continued service and representative of the overall health.
- The index should contain objective and verifiable measures of different underlying asset condition at the chosen hierarchical level, as opposed to subjective observations.
- Consolidate all different asset condition in a single integrated view of aggregated HI which is understandable and readily interpreted.

While the definition of aggregation can vary based on the application, we aim to define aggregated HI as an indicator that must represent, at best, the health state at the desired level (bay or substation level) based on the HI score of individual assets. In this way, the aggregation will take into account the multiple modes of degradation and failure modes that characterizes an asset. Depending on the index and its interpretation, the asset managers will be able to make decisions keeping in mind that asset management process also factors in risk levels and other corporate priorities which may override purely condition-based decisions. Moreover, such an will be able to answer if certain geographic regions wear out asset faster. Clear data trends and facts in an aggregated manner will form a more solid basis for discussions on how to mitigate risks.

### *B. Benefits of aggregated HI for better decision making*

Aggregation will help in better decision making process at any desired hierarchical level by serving as a single global indicator that will represent a snapshot of the overall health considering the health of underlying individual assets. Fig. 2 shows how increasing level of aggregation is related to the details. For example, at the bay level within a substation, this can be achieved by modeling the importance of condition (or similar) indicators, failure modes or similar chosen reference and the current health condition of individual assets. It will play a pivotal role to address the three key challenges faced by utilities:

*1) Ageing infrastructure and maintenance strategy improvement*

In the majority of the European and in developed countries, it is evident that a large portion of infrastructure assets are nearing their end-of-life or require significant maintenance investments to maintain consistent levels of reliability [10]. For any utility, assets are valuable and capital intensive. Moreover, asset failures near its remaining useful life tend to be more expensive than preventive refurbishment or replacement, as failures occur at inconvenient times, require service restoration action and temporary measures [11]. Ageing assets are causing an increasing number of *emergency* maintenance interventions, which are both cost-extensive as well as harder to manage.

*2) Lack of resources and corrective actions prioritization*

With the ageing infrastructure, there is also ageing workforce as well as lack of right resources to execute the tasks and it poses a serious concern for many utilities [12]. An ageing workforce means implicit knowledge is leaving the organizations. Thus, moving towards a data-driven decision making using different aggregation techniques will help to capture the expert knowledge before it leaves the organization.

*3) Budget and maintenance expenses*

Asset managers always tend to build up resource pressure to keep costs to a minimum, i.e., a right balance between CapEx and OpEx. Ageing assets require investment for their maintenance. It might happen that utilities and regulators can demand a detailed analysis and justification for CapEx and OpEx budgets being submitted for refurbishments and/or replacements. At any chosen hierarchical level, aggregated HI help the utilities to balance these competing factors by providing insights about underlying asset health and condition. In addition, aggregated HI provides long term views about necessary maintenance activities and system reliability.

The above mentioned benefits of HI aggregation will help in tackling the perfect storm that is brewing on the ageing grid infrastructure. In response to the increasingly demanding requirements in terms of maintenance interventions and other asset management activities, the application of aggregated HI offers the possibility to improve the process of decision making in maintenance planning.

## III. Different HI Aggregation Techniques

### A. *Different techniques of aggregation*

Aggregated HI, by their nature, will indicate the condition of the respective bay or substation or any chosen hierarchical level and whether any asset should be repaired or replaced. Choosing an aggregation technique depends on the requirements of asset managers or any other user. For example, aggregation can be performed at bay level where the aggregated HI will give insights into different assets in the same bay. Or the aggregation can be performed for asset type in the substation, say all power transformers in one substation. Comparing the HI for multiple transformers with the aggregated HI will reveal insights and trends that were either not previously evident, or based on expert opinion from experienced field engineers, but not verified with facts. Other possible factors that can be considered are:

- Redundancy of asset in chosen hierarchy
- Costs of asset replacement
- Time to replace the asset
- Spare asset availability
- Operational stress
- Age of bay

During the exploration of the possibilities for HI aggregation, the following four methods were determined to calculate the aggregated HI score:

1. Use an average weight with rules from expert opinion
2. Use Failure Modes, Effects and Criticality Analysis (FMECA)
3. Use replacement cost as weighing factor
4. Use HI scoring with failure interpretation

TABLE I. Average Weight (#1)

| Advantages | Disadvantages |
|---|---|
| • Equal weights are easy to use<br>• Extra rules make sure that no bad scoring gets lost<br>• Universal method to calculate the HI of a bay (does not matter which HI score is used)<br>• Results are also available per group<br>• Easy to program/create/implement | • Importance of asset is not used for aggregation<br>• Extra rule is arbitrary<br>• HI score is needed to calculate the HI of a bay |

TABLE II. FMECA (#2)

| Advantages | Disadvantages |
|---|---|
| • Weighing based on the influence of each asset to the parent system<br>• Use of the current decision that are already in use | • FMECA in good quality is needed<br>• Criticality is already added to the weighing factors<br>• Need to restructure the FMECA to fit the calculation |

TABLE III. Replacement Cost as Weighing Factor (#3)

| Advantages | Disadvantages |
|---|---|
| • Low HI can indicate the weakness of system in a simple way<br>• Easy to interpret, i.e., weighting factor can be changed on demand<br>• Weighing based on logarithmic scales that highlights poor asset conditions | • Along with technical HI, if more HI's (operation, financial) are added, then this cannot be compared to the other methods directly<br>• Two HI need to be determined for each system level. Weighting factor of purchasing costs need to be determined for each type. Weighting factor with purchasing costs can only give the indication of the general condition, but cannot prioritize the decision<br>• Asset HI is needed for calculation |

TABLE IV. HI scoring with Failure Interpretation (#4)

| Advantages | Disadvantages |
|---|---|
| • Assigns a score which has a defined meaning | • Does not forecast the expected financial burdens of replacements<br>• Can only be applied if asset HI scores are calculated the same way everywhere<br>• Requires assessment of each asset's ability to cause its parent system to fail<br>• Does require a score definition for each system |

### B. *FMECA based severity*

In this study, we take a closer look at the method #2. Failure Modes, Effects and Criticality Analysis (FMECA) is a technique to break a system down into its elements to explore the possible failure modes and its corresponding effects. Thereafter, criticality analysis is performed to prioritize the failure modes for potential treatment [13]. Analysing failure modes and conducting root cause analysis will provide insight into the conditions leading up to asset failures. Once the factors leading to failures are known, assets can be monitored for those factors and repaired prior to failure. FMECA helps in classifying the large population of assets and to decide which assets require more attention and what actions should be taken. The complexity of such decision-making increases because each asset class has different failure modes, and each failure has different consequences in the bay and eventually in the substation [14]. FMECA is widely accepted within utilities for failure investigation and we use the methodology to convince the practical aspects. To calculate the weights for aggregation, FMECAs are used. With the FMECA-based method every asset gets a weight according to their importance in the parental system.

## IV. Case Study

In this study, a hierarchy approach (graded/ranked) series is followed incorporated by a bottom-up approach (from asset towards bay). Fig. 3 illustrates the hierarchy of HI calculation which starts from asset level to bay level and up to station level. Such an approach can be easily extended to other structures wherever a hierarchy can be built up. The aim is to have a tree-kind structure where the $HI_{bay}$ is a summation of $HI_{all}$ individual assets in that specific bay (referring to Fig. 3). There is always a possibility of moving upwards in the hierarchy.

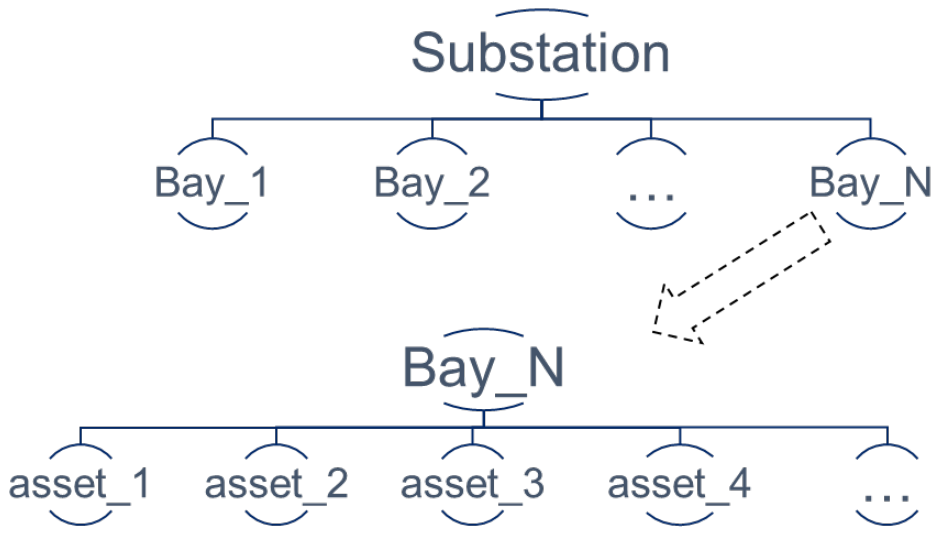

Figure 3. Hierarchical approach from substation to bay to individual assets

Based on expert opinion within TenneT, this study aims to compare the methods #1 and #2. To perform the comparative analysis, 5 substations (mix of small and big) from TenneT are chosen. In this way, both methods could be validated on real data. The substations are coded for confidentiality. HI color code and score followed at TenneT is used and is shown in Table V(A) [15]. The inputs to the methods are data from different sources related to the health and FMECA of the asset. For method #1, based on expert opinion, the whole asset population is divided into primary, secondary, and tertiary assets. The weightage for each asset type is based on detailed assessment by experts. It is then used to normalize the values, weighting them with corresponding weights and building a personalized average scoring. Table V(B) shows the average scores used for method #1. The aggregated HI is defined as:

$$\sum \begin{Bmatrix} HI_{CB} \\ HI_{disc} \\ HI_{...} \end{Bmatrix} \times \begin{Bmatrix} W_{CB} \\ W_{disc} \\ W_{...} \end{Bmatrix} = HI_{Bay}$$

where, $HI_{bay}$ is a summation of $HI_{\text{all individual assets}}$ in that specific bay. Since bay sizes (in terms of number of assets) can vary, the cumulative HI score will vary as well. However, it is normalized to a nominal range of 1-10.

For method #2, the asset severity is calculated from FMECA with the help of experts and is shown in Table VI. The aggregated HI in method #2 is defined as:

$$HI_{i,agg} = \frac{HI_i * \sum_{j=1}^{J} S_{i,j}}{\sum_{j=1}^{J} \sum_{i=1}^{I} S_{i,j}} \qquad (1)$$

where, i = no. of assets, j = no. of severity from FMECA, N = no. of assets, and S = Severity.

Table VII displays the comparative study of methods #1 and #2 for the 5 substations. The 5 substations have assets with quite varied HI scores. While the pie-chart shows the distribution of assets based on HI, the bar-plots show the bay HI based on methods #1 and #2. Surprisingly, substation VL reports 18.2% assets with HI score of 0. This is a clear sign of bad data, and it affects the whole organization. While it was possible to clean the data before analysis, it was intentional to use the data as it is. The reason is to showcase the result biasing based on bad or invalid data. It is a fundamental assumption for any data aggregation methodology intended for real world application. And, it must be built around the fact that the data to be gathered will be most certainly incomplete and to a certain extent inaccurate (or in some cases contradictory).

For the comparative analysis, it is noticeable that method #1 biases the aggregation method. This is clear when we look at the bays ZYRA, ZYRB and ZYBS in the substation ZY. A closer look reveals that those bays are composed of few assets resulting in low aggregation score. On the other hand, use of FMECA method changes the result completely. Unfortunately, the full explanation of all substations is out of scope of the present paper and will be fully treated in a separate detailed study.

TABLE V. (A) HI COLOR CODE WITH CORRESPONSING SCORE, (B) ASSET POPULATION TYPE AND SCORE (#1)

| HI_Color |
|---|
| Green (HI_score 9-10) |
| Orange (HI_score 7-8) |
| Red (HI_score 4-6) |
| Violet (HI_score 1-3) |
| White (HI_score 0) |

(A)

| Asset population type | score |
|---|---|
| Primary | 7 – 10 |
| Secondary | 4 – 6 |
| Tertiary | 1 – 3 |

(B)

TABLE VI. ASSET AND SEVERITY FROM FMECA LIST OF FAILURES (#2)

| Type of asset | FMECA severity |
|---|---|
| Earthing | 343 |
| Compensation coil | 304 |
| Protection devices | 152 (built before 1992)<br>237 (built after 1992) |
| Power transformer | 458 |
| Overvoltage Surge Arrestor | 128 |
| Disconnector | 313 |
| Instrument Transformer | 377 |
| Control device | 148 |
| Circuit breaker | 464 |
| Power transformer | 458 |

TABLE VII. COMPARATIVE STUDY OF METHODS #1 AND #2 (X-AXIS: BAY, Y-AXIS: HI SCORE)

**Station:** DI
**No. of bays:** 21
**No. of assets:** 557

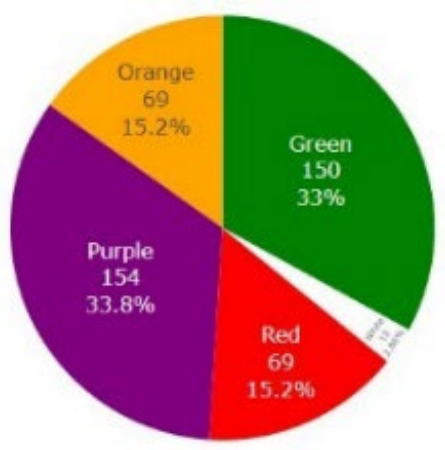

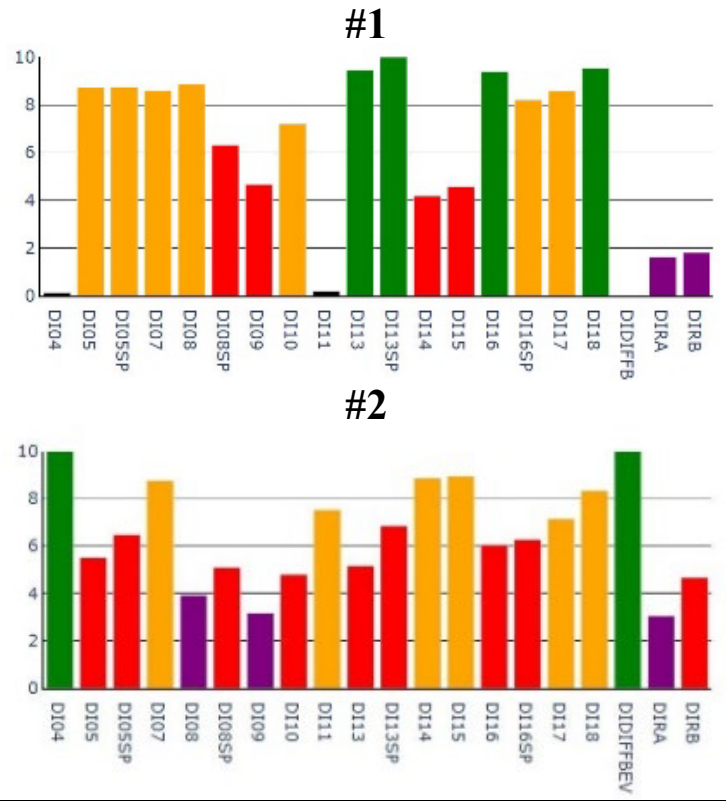

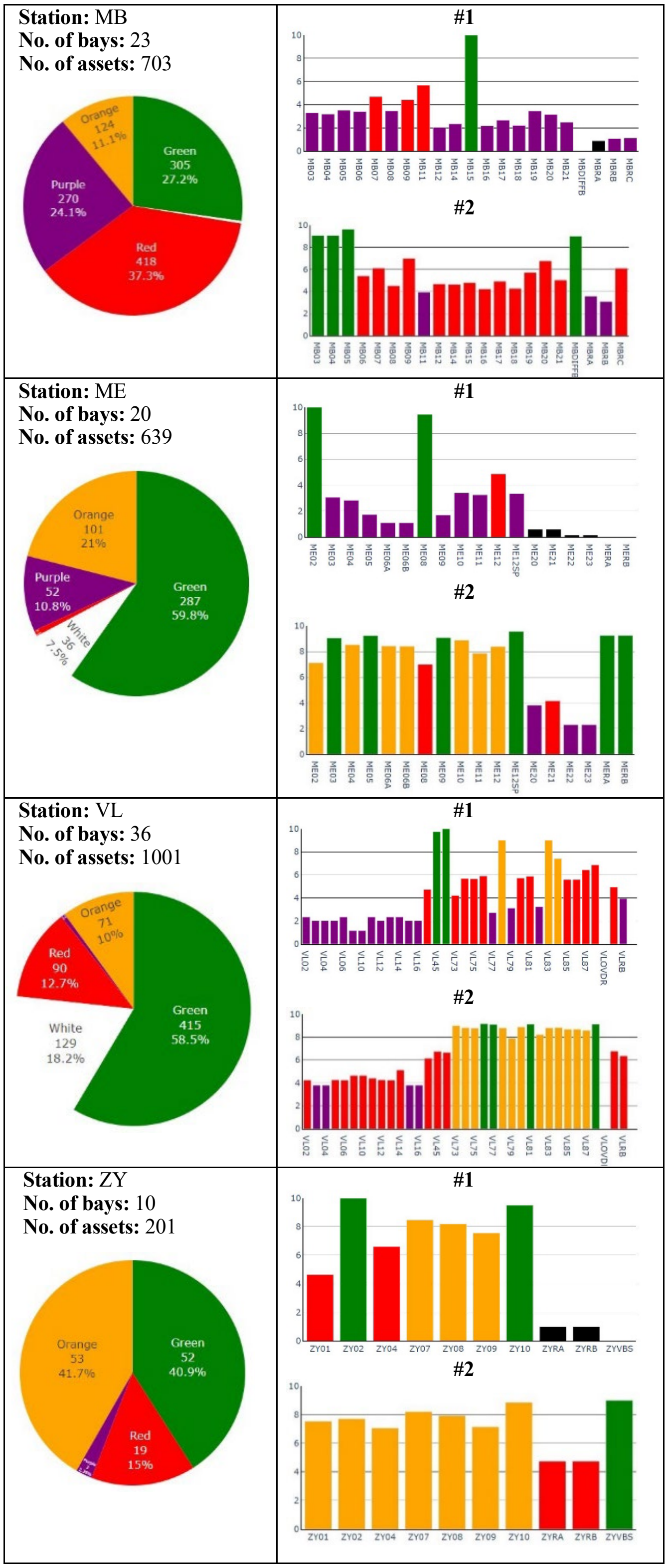

## V. Conclusion

The building of a single indicator, representing the system health, was formulated in this study. Such a system health indicator at bay level was based on the aggregation of HI of several assets and their severity from FMECA. However, the downside of such an approach was the loss of transparency and supporting detail such as the relative contributions of major asset groups, or of cost categories such as deterioration versus consequence and likelihood of failure. Several benefits and challenges to HI aggregation were also presented in this study. The benefits of using an aggregated HI score is its conciseness and simplicity, coupled with the ability to aggregate information about multiple asset groups and condition metrics into a single value. Some of the challenges include invalid data, choice among different methodologies, and undefined data model. With this study, it is envisioned that the concept of aggregated HI will help asset managers to drive maintenance and replacement strategies as well as consistently compare the system health at the bay level (or similarly different hierarchical level).